\documentclass{llncs}

\usepackage{cmap}

\usepackage[utf8]{inputenc}
\usepackage{soulutf8}
\usepackage{color}
\usepackage[autostyle]{csquotes}

\usepackage[bookmarks=false,breaklinks,hidelinks]{hyperref}
\hypersetup{hidelinks,
   colorlinks=true,
   allcolors=black,
   pdfstartview=Fit,
   breaklinks=true}

\usepackage{multicol}
\usepackage{listings}
\usepackage{breakurl}
\usepackage{multirow}
\usepackage{booktabs}
\usepackage{makeidx} 
\usepackage{graphicx}
\usepackage{caption}
\usepackage{subcaption}
\usepackage{paralist}
\usepackage{soul}
\usepackage{cite}
\usepackage{url}

\usepackage{color}
\usepackage{xcolor}
\usepackage{pdfcomment}

\usepackage[all]{hypcap}

\definecolor{codegreen}{rgb}{0,0.6,0}
\definecolor{codegray}{rgb}{0.5,0.5,0.5}
\definecolor{codepurple}{rgb}{0.58,0,0.82}
\definecolor{backcolour}{rgb}{0.95,0.95,0.92}

\lstdefinestyle{mystyle}{
    backgroundcolor=\color{backcolour},
    commentstyle=\color{codegreen},
    keywordstyle=\color{magenta},
    numberstyle=\tiny\color{codegray},
    stringstyle=\color{codepurple},
    basicstyle=\fontsize{7}{9}\ttfamily,
    breakatwhitespace=true,
    breaklines=true,
    captionpos=b,
    keepspaces=true,
    numbersep=5pt,
    showspaces=false,
    showstringspaces=false,
    showtabs=false,
    tabsize=4,
    frame=tblr,
    framerule=0pt,
    columns=flexible
}
\colorlet{punct}{red!60!black}
\definecolor{background}{HTML}{EEEEEE}
\definecolor{delim}{RGB}{20,105,176}
\colorlet{numb}{magenta!60!black}

\lstdefinelanguage{json}{
    basicstyle=\normalfont\ttfamily,
    numbers=left,
    numberstyle=\scriptsize,
    stepnumber=1,
    numbersep=8pt,
    showstringspaces=false,
    breaklines=true,
    frame=lines,
    backgroundcolor=\color{background},
    literate=
     *{0}{{{\color{numb}0}}}{1}
      {1}{{{\color{numb}1}}}{1}
      {2}{{{\color{numb}2}}}{1}
      {3}{{{\color{numb}3}}}{1}
      {4}{{{\color{numb}4}}}{1}
      {5}{{{\color{numb}5}}}{1}
      {6}{{{\color{numb}6}}}{1}
      {7}{{{\color{numb}7}}}{1}
      {8}{{{\color{numb}8}}}{1}
      {9}{{{\color{numb}9}}}{1}
      {:}{{{\color{punct}{:}}}}{1}
      {,}{{{\color{punct}{,}}}}{1}
      {\{}{{{\color{delim}{\{}}}}{1}
      {\}}{{{\color{delim}{\}}}}}{1}
      {[}{{{\color{delim}{[}}}}{1}
      {]}{{{\color{delim}{]}}}}{1},
}

\usepackage{amsmath}
\usepackage[capitalise,nameinlink]{cleveref} 
\crefname{section}{Sect.}{Sect.}
\Crefname{section}{Section}{Sections}

\newcommand{\classname}[1]{\texttt{#1}}

\begin{document}

\title{A Case Study on Visualizing Large Spatial Datasets in a Web-based Map Viewer\thanks{This work has been funded by 
Xunta de Galicia/FEDER-UE CSI: ED431G/01; GRC: ED431C 2017/58. 
MINECO-CDTI/FEDER-UE CIEN LPS-BIGGER: IDI-20141259; INNTERCONECTA uForest: ITC-20161074.
MINECO-AEI/FEDER-UE Datos 4.0: TIN2016-78011-C4-1-R; Flatcity: TIN2016-77158-C4-3-R.
EU H2020 MSCA RISE BIRDS: 690941.}}

\author{Alejandro Corti\~nas\orcidID{0000-0002-2555-6342} \and 
Miguel R. Luaces\orcidID{0000-0003-0549-2000} \and 
Tirso V. Rodeiro\orcidID{0000-0003-2373-0746}}

\institute{Universidade da Coru\~na\\
Laboratorio de Bases de Datos\\
A Coru\~na, Spain\\
\email{\{alejandro.cortinas, luaces, tirso.varela.rodeiro\}@udc.es}
}

\maketitle

\begin{abstract}
Lately, many companies are using Mobile Workforce Management technologies combined with information collected by sensors from mobile devices in order to improve their business processes. Even for small companies, the information that needs to be handled grows at a high rate, and most of the data collected have a geographic dimension. Being able to visualize this data in real-time within a map viewer is a very important deal for these companies. In this paper we focus on this topic, presenting a case study on visualizing large spatial datasets. Particularly, since most of the Mobile Workforce Management software is web-based, we propose a solution suitable for this environment.
  
\end{abstract}

\section{Motivation}\label{intro}

Mobile Workforce Management (MWM) technologies are increasingly being used by companies to manage and optimize their workers' task schedules and to improve the performance of their business processes~\cite{oracle2014}. These technologies, used in combination with the information collected by current mobile technology (e.g. the geographic position using a GPS receiver, or the user activity using an accelerometer), are useful to detect patterns in the past activity of workers, or to predict trends that can improve the future scheduling.

Datasets produced by mobile sensing and MWM technologies are large and complex. As an example, consider a small package delivery company with a fleet of 100 vehicles, each one producing a GPS position every 10 seconds (64 bytes taking into account a device id, a timestamp, three geographic coordinates, speed, bearing, and accuracy). Supposing that each vehicle is active 8 hours per day, each one would produce 2,880 events generating 184,320 bytes of data every day, and the company would require over 17 MB of storage per day. Larger systems (e.g., MRW, a Spanish package delivery company, declares to have more than 3,300 vehicles), or the inclusion of additional sensor data (such as accelerometer data) would produce larger datasets.

MWM technologies often require web-based dashboards to visualize and query the information stored in the system. Moreover, given that the information is of geographic nature, these dashboards require GIS technology such as map server and map viewers. Nah cites in~\cite{nah2004study} a number of studies that propose that web users accept waiting between 1 and 42 seconds for a web page to load, but it concludes that, considering purposeful browsing for information retrieval tasks as opposed to open-browsing, most users are willing to wait for only about two seconds. Even though the study considers that users are browsing the web and not using a web-based dashboard, we believe that a waiting interval of two seconds for a page refresh is sensible.

Data management technologies have been working during the last years to support horizontal scaling and distributed processing. Hence, storing and querying large geographic datasets can be achieved using different technologies. However, choosing the most appropriate technology to support these usage scenarios is a complex task. Furthermore, current web-based GIS technology are not designed to achieve browsing of large datasets with a latency of less than 2 seconds. For example, middleware software such as map servers have little support for NoSQL technologies, and visualization software such as map viewers aggregate geographic information on the client side, thus requiring large datasets to be transferred over the network and to be processed in the web browser. Hence, in order to support the visualization of large geographic datasets, middleware components and map viewers must support querying and aggregating geographic data using distributed processing systems.

In this paper, we present a case study on visualizing large spatial datasets in a web-based map viewer. We aim at identifying the most suitable technology, proposing an alternative to achieve data visualization with a latency smaller than two seconds. In~\cref{background} we describe our previous work and the system architecture that we propose. In~\cref{background} we present the research questions that we want to answer with the case study and the evaluation methodology. In~\cref{experiments} we show the experiments that we have performed and the results we have achieved. Finally, in~\cref{conclusions} we present our conclusions and future work.
\section{Previous Work and System Architecture}\label{background}

\begin{figure}[tbp]
 \centering
 \includegraphics[width=\textwidth]{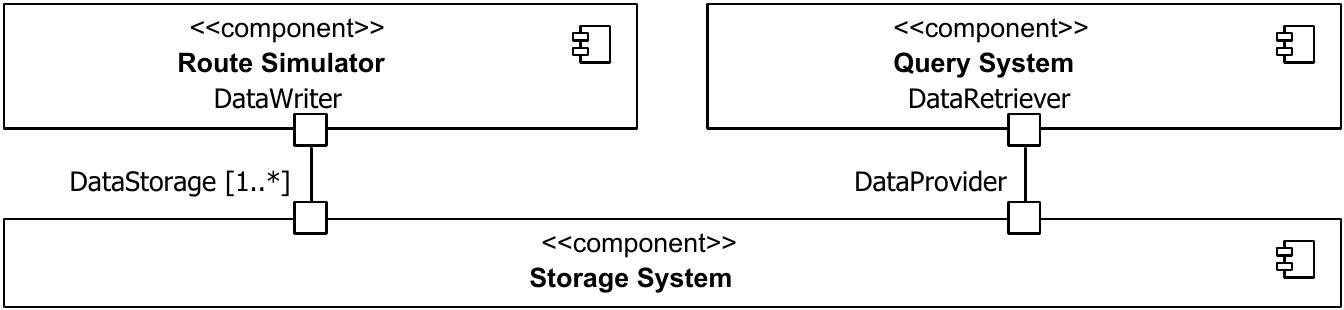}
 \caption{System architecture}
\label{general-arch}
\end{figure}

We have presented in a previous paper~\cite{cortinas2018} the architecture of a system to store, query and visualize on the web large datasets of geographic information (see~\cref{general-arch}). The architecture includes a component to simulate a large number of drivers that circulate through a road network and report their position to the server on a regular basis (\classname{Route Simulator}). In addition, the architecture provides a \classname{Storage System} with exchangeable storage subsystems so that they can be tested under the same load conditions and evaluate their performance with the same queries. \cref{ingestion-event} shows an example of an event received and stored by the system. It consists of the driver id, the GPS position of the worker, the timestamp of the position, and additional information in JSON format that is specific of the particular domain for which the architecture is being used. Finally, the architecture also includes a component to solve queries and cluster data that is visualized in a web-based map viewer (\classname{Query System}).

\begin{figure}[tbp]
\begin{lstlisting}[language=json,numbers=left]
{
 "driver_id"; 3,
 "position": {
    "x": -4.013856742, "y": 40.358347874, "z": 517,
    "speed": 32.48,
    "bearing": 83.6,
    "accuracy": 4.5509996
  },
  "timestamp": 1513763866,
  "data": {...additional data in json format...}
 }
\end{lstlisting}
\caption{Example of an event received by the \classname{Storage System} component}
\vspace{-3ex}
\label{ingestion-event}
\end{figure}

\begin{figure}[tbp]
 \centering
 \includegraphics[width=\textwidth]{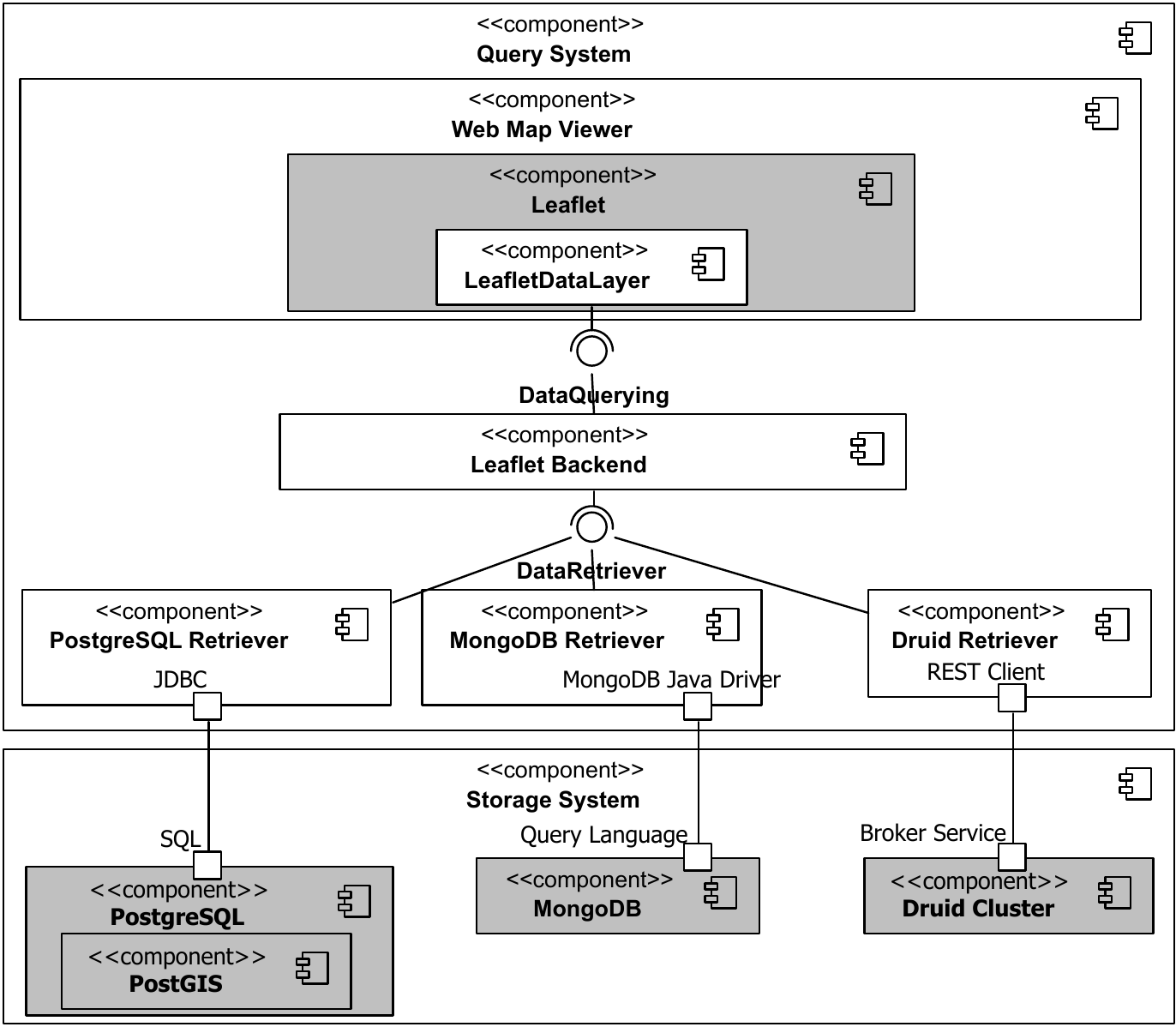}
 \caption{Detailed querying architecture}
\label{querying-arch}
\end{figure}

Figure \ref{querying-arch} shows a detailed view of the querying architecture components. The components with a gray background are third-party components that are used without modifications. The communication with the \classname{Storage System} component is managed by a component that implements the generic \classname{DataRetriever} interface. We have currently implemented three alternatives: one that retrieves the events from Postgres\footnote{https://www.postgresql.org/} + PostGIS\footnote{https://postgis.net/} (the component \classname{PostgreSQL Retriever}), another one that retrieves the data from MongoDB\footnote{https://www.mongodb.com} (the component \classname{MongoDB Retriever}), and another that retrieves the data from Druid\footnote{http://druid.io/}~\cite{Yang:2014:DRA:2588555.2595631} (the component \classname{Druid Retriever}). Queries are sent from a \classname{Web Map Viewer} component, implemented using \classname{Leaflet}, by a client-side component called \classname{LeafletDataLayer} implementing the \classname{Layer} interface of Leaflet. A server-side component called \classname{Leaflet Backend} receives the queries, delegates them to the appropriate data retrieving component, and sends back the results to the client-side.

Considering that having a fluid visualization in the client side is a very important requirement, we have to aggregate the points on the server side of the application and send to the client side only the result of the aggregation instead of transferring large collections of individual geographic points to be aggregated on the client-side. Furthermore, considering that the user will define specific spatial and temporal ranges for the set of events that have to be retrieved by means of zoom and pan operations in a map and a time range control, precomputed clusters cannot be used because the variation among queries is too large. The simplest alternative is to perform the query \emph{get all points in the range (xmin, ymin, tmin) - (xmax, ymax, tmax)} and apply a clustering algorithm on the result, but it is a costly solution in terms of computation requirements. Instead, taking into account that in a geographic reference system where the coordinates represent longitude and latitude in degrees a value with an accuracy of 9 decimals represents a maximum of 1 millimeter on the surface of the Earth, in~\cite{cortinas2018} we proposed to store 7 additional versions of the same geographical point with 7 different precisions (between 2 and 8 decimals). This makes the process of clustering  as simple as grouping the events by equal values of coordinates and counting the number of elements. Moreover, computing additional versions of each geographic point is assumable in storage cost and insertion time.

Our tests in~\cite{cortinas2018} revealed that this approach cannot be used to achieve a constant time in aggregation queries because truncating a decimal means that one point in a level of aggregation represents one hundred points in the next level of aggregation. Thus, the difference between the different levels of aggregation is too high. Furthermore, the aggregated data required another brief aggregation step in the client side in order to draw the different aggregated elements to make the map look nice to the user. Hence, we decided to follow a different approach to determine the aggregation levels taking into account the final visualization.

When a user navigates in a web map viewer, it sends queries to the server depending on the current view to retrieve the data that has to be shown. Each of these queries is associated with the bounding box of the current view, that is, the maximum and minimum latitude and longitude of the view. Regarding aggregation, there is another parameter that affects the actual representation of the aggregated elements in the map viewer: the zoom or scale of the current view. If we are seeing the map with very little zoom, the map viewer needs to aggregate more in order to show a suitable view, and the other way around. 

We propose to compute discretized versions of the geographic points according to the zoom level. We consider 18 different zoom levels, which is common in GIS visualization, so we store, for each point, 18 alternative versions of it. For each level of zoom, the separation between aggregated elements is calculated using Formula~\ref{eq:1}, which maintains the same distance for the aggregated elements independently of the zoom level in a map viewer. For example, for the zoom level 11, the aggregated elements should be separated 0.043945312 degrees\footnote{We are not considering different separations for latitude and longitude because for our case study location, Spain, using the same separation for both is adequate.}. When we store the alternate version of a point for the zoom level 11, we calculate the closest multiple of 0.043945312 to both the latitude and the longitude of our point. All the points that need to be aggregated at zoom level 11 have the same alternative location. Obviously this approach makes each point to take much more space, but since we are focusing on fluid visualization, we assume this drawback. 

\begin{equation}\label{eq:1}
  separation(zoom)=\begin{cases}
    90º, & \text{if $zoom=0$}\\
    separation(zoom - 1)/2, & \text{if $zoom>0$}\\
  \end{cases}
\end{equation}

\section{Case Study Objectives}\label{objectives}

In order to validate the architecture of the system proposed in~\cref{background}, we have identified two key aspects that have to be evaluated described in the following research questions:

\begin{itemize}
    \item \emph{Research question 1 (RQ1). Can we build a web-based map viewer for large geographic datasets with a latency lower than 2 seconds in data refreshes?} This research question will test whether we can use the simple approach proposed in~\cref{background} to improve the response time of aggregation queries.
    
    \item \emph{Research question 2 (RQ2). Which of the candidate storage technologies provides a faster answer to aggregation queries?} Even though the selection of a storage technology must take into account many requirements (e.g., transaction support, horizontal scaling, etc.), being able to answer aggregation queries is a very important requirement in our architecture.
\end{itemize}    

\begin{figure}[tbp]
\centering
\begin{subfigure}[t]{.5\textwidth}
  \centering
     \includegraphics[width=\linewidth]{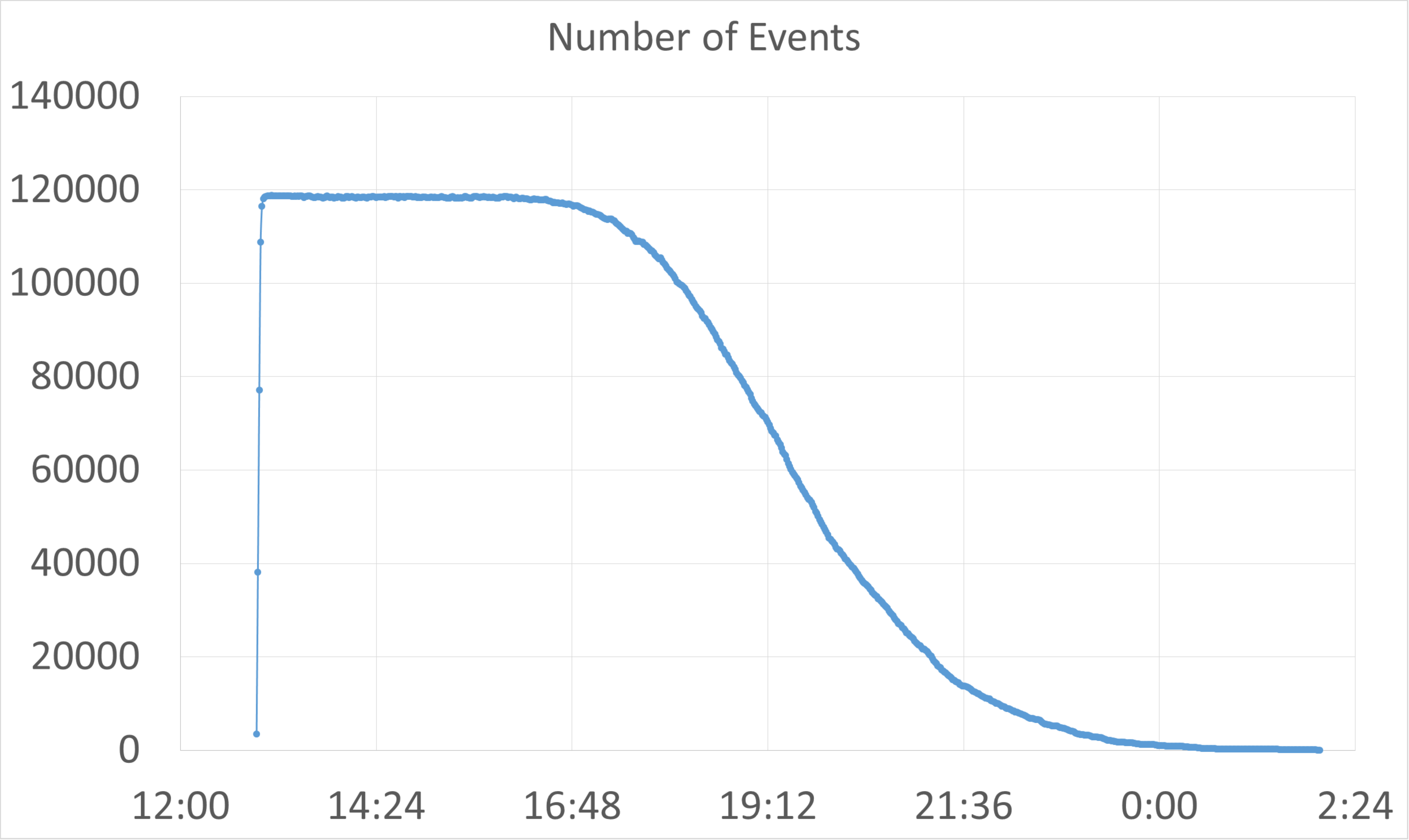}
  \subcaption{Data distribution over time}
  \label{fig:data-time}
\end{subfigure}%
\begin{subfigure}[t]{.5\textwidth}
  \centering
	\includegraphics[width=0.98\linewidth]{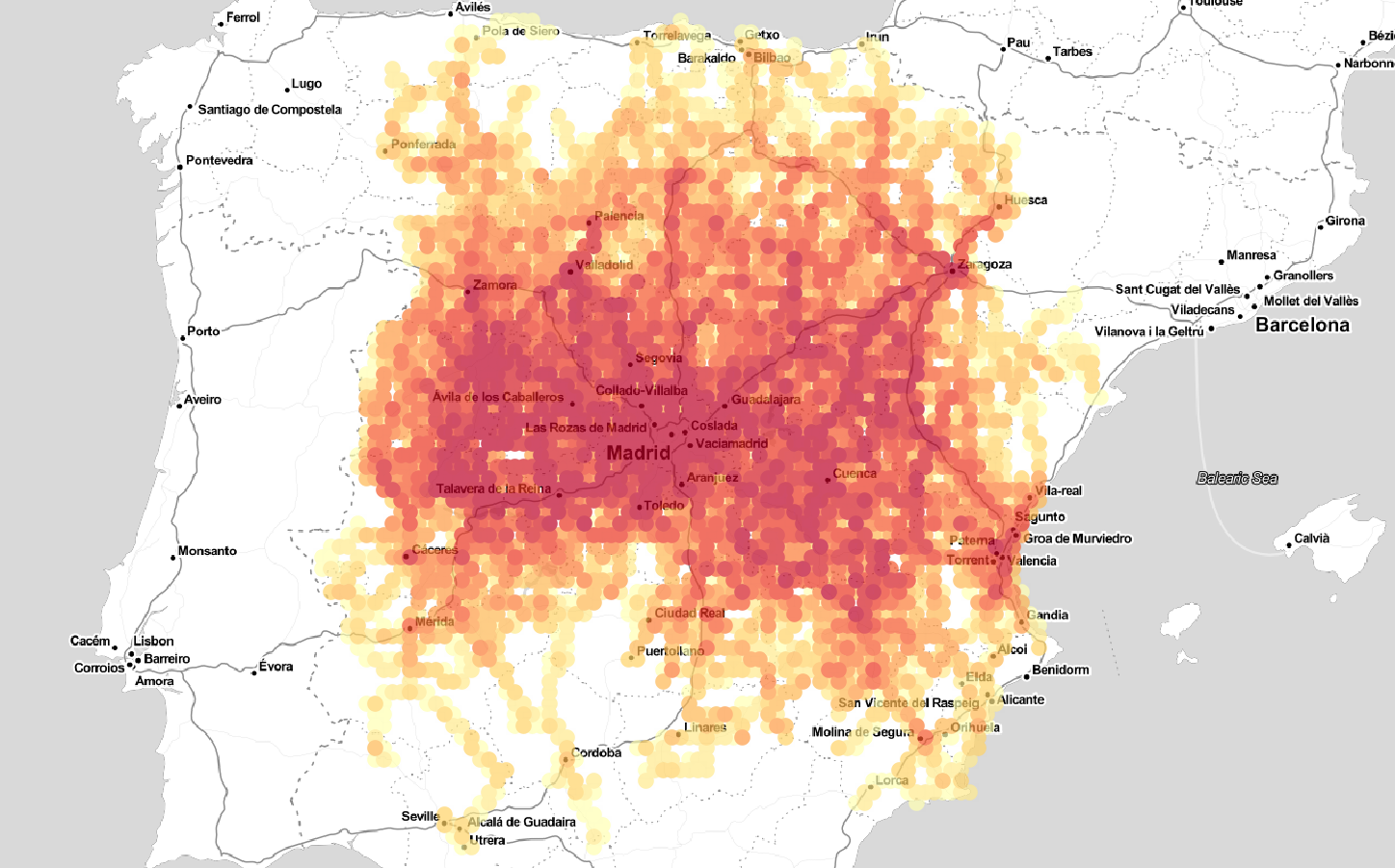}
   \subcaption{Data distribution over space}
  \label{fig:data-space}
\end{subfigure}
\label{fig:data}
\vspace{-1ex}
\caption{Data distribution over time and space}
\vspace{-2ex}
\end{figure}

To evaluate these research questions, we have run the \classname{Route Simulator} component in a desktop computer (Intel Core i7-3770, 4 cores, 3.40GHz, 8GB of RAM) to generate events for 2000 simultaneous drivers driving for approximately 14 hours, resulting in a dataset of 47.8 million events. Each driver starts at a random position in the road network, it computes a route to a random destination, and it generates positions along the sections of the route every second assuming a random speed expressed as a percentage of the maximum allowed speed. For example, a driver can circulate at 80 \% of the maximum speed of a road segment, and then circulate at 105 \% of the maximum speed in the next road segment. In the \classname{Route Simulator} component, we have separated the dataset generation step from the dataset ingestion step in order to ensure that exactly the same datasets are stored in each storage technology. \cref{fig:data-time} and \cref{fig:data-space} shows the data distribution over time and space. The distribution over time shows that all drivers start simultaneously and finish smoothly. The distribution over space shows that the positions are distributed following the population density.

In order to evaluate RQ1 and RQ2, one hundred queries were randomly generated with six different levels of zoom, from a higher zoom level (15) to a lower one (10). 
To generate realistic queries with different zoom levels, we used the same map viewer (same width and height), to calculate the spatial ranges. 
Therefore a higher zoom level represents a smaller spatial range query, and the other way around. For each zoom level, the alternative version stored for it was used for the aggregation as described in \cref{background}. Each query was executed exactly once to avoid the effects of any possible caching. 

\section{Experiments and Results}\label{experiments}

The experiments were run on a machine (Intel Core i5-4440, 4 cores, 3.10GHz, 16 GB of RAM) that hosted all the server-side components of the architecture (\classname{Storage System}, and \classname{Query System}). Only one storage technology was running simultaneously (either Postgres+PostGIS, MongoDB or Druid) in order to avoid resource allocation competitions. 

\begin{figure}[tbp]
\centering
\includegraphics[width=\linewidth]{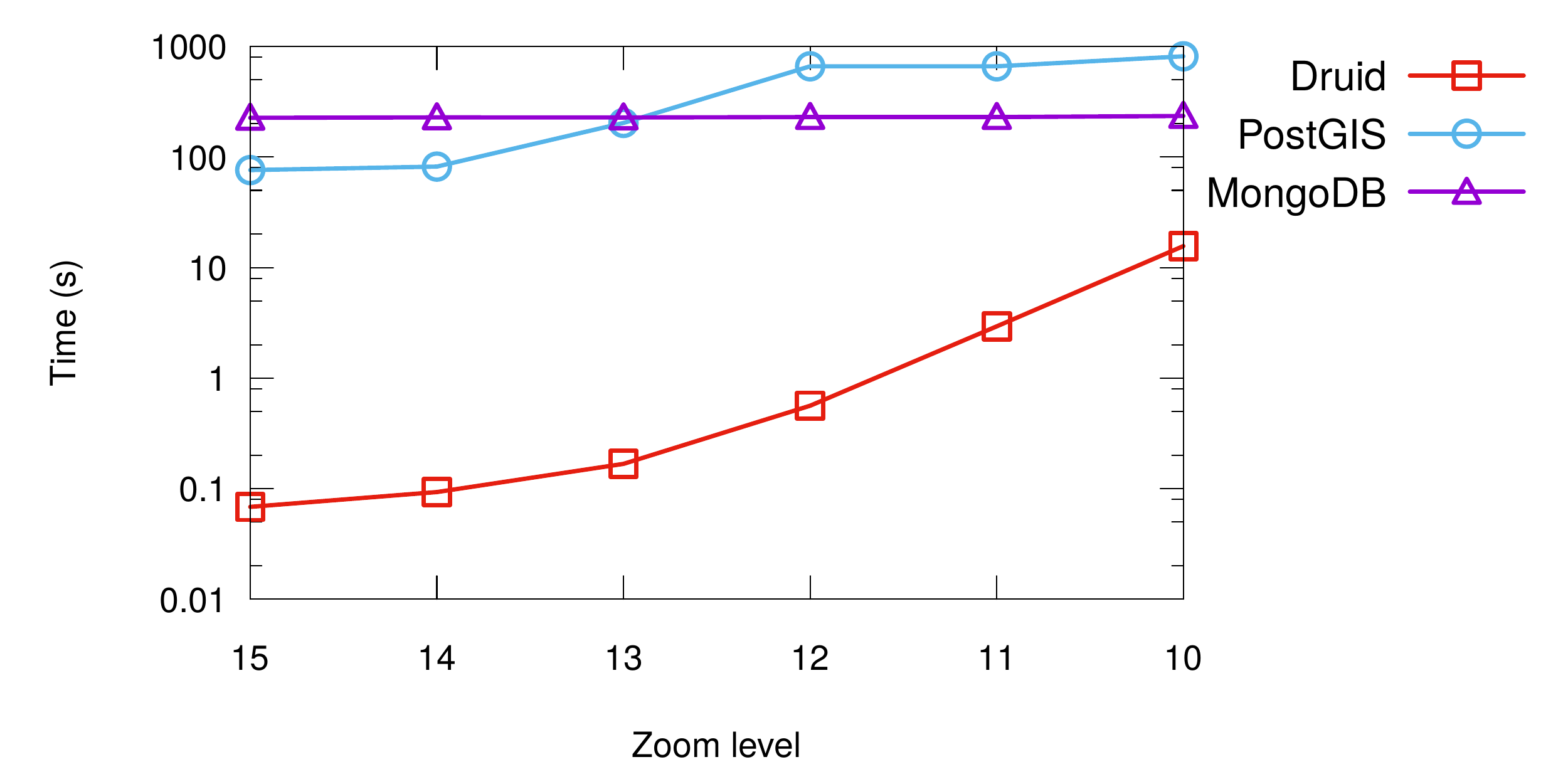}
\vspace{-2ex}
\caption{Results of the experiments}
\label{figure:results}
\vspace{-2ex}
\end{figure}

\Cref{figure:results} show the results to evaluate RQ1 and RQ2. The horizontal axis represents different zoom levels from high (smaller spatial range queries) to low (bigger spatial range queries). The vertical axis represents the average time in seconds to answer 100 queries using a logarithmic scale. We can see that Druid is the only technology able to resolve queries below 2-3 seconds, but only when the zoom level is 11 or higher. Particularly, when the zoom level is 11 the average query time is 2.93847 seconds. In our previous work, Postgres+PostGIS results were close to the Druid ones, but it seems that the extra size required to store the 18 alternative versions or the higher density of events totally invalidated Postgres+PostGIS in this case study. 

The results obtained indicate that with our approach we can build a web-based map viewer for large geographic datasets with a latency close or lower to 2 seconds but only in certain conditions (zoom level below 11, approximately a width of 40 km in a 600 px wide map viewer). Zoom levels above 11 imply retrieving extremely large collections of geographic points, and the only suitable approach seems to be precomputing estimations for the clusters. This conclusion was also validated using a web-based map viewer to visualize the evaluation dataset. The results also determine that Druid is the best option for the storage technology, matching the conclusion from our previous work. 

\section{Conclusions}\label{conclusions}

We have presented in this paper a case study of a web-based map viewer for large geographic datasets with a latency close or lower to 2 seconds. The case study has shown that storing additional versions of each geographic point and using a columnar database designed to answer OLAP queries can be used to achieve this goal. The case study was also designed to help selecting the best technology to store and query large geographic datasets. Whereas our previous research showed that PostgreSQL+PostGIS was comparable to Druid in terms of efficiency, the extended dataset that we generated this time shows that PostgreSQL+PostGIS performs worse than Druid. The source code for our experiments can be found at the research group GitLab\footnote{https://gitlab.lbd.org.es/groups/massive-geo-data}.

As future work, we need to compare these results with many other technologies such as cstore\_fdw\footnote{https://citusdata.github.io/cstore\_fdw/}, a PostgreSQL columnar extension, or NoSQL technologies oriented to store time-series such as InfluxDB\footnote{https://www.influxdata.com/}. Since we cannot find a technology to solve large spatial range queries in an acceptable time, we are also working on designing a data structure able to resolve these kind of queries. A future line of work is also testing the influence of the temporal dimension to these queries.  

\bibliographystyle{splncs03}
\bibliography{library}

\begin{thebibliography}{1}
\providecommand{\url}[1]{\texttt{#1}}
\providecommand{\urlprefix}{URL }

\bibitem{cortinas2018}
Corti{\~{n}}as, A., Luaces, M.R., Rodeiro, T.V.: {Storing and Clustering Large
  Spatial Datasets Using Big Data Technologies}. In: Proceedings of the 16th
  International Symposium on Web and Wireless Geographical Information System
  (W2GIS 2018). A Coru{\~{n}}a (2018), (Pending publication)

\bibitem{oracle2014}
Creelman, D.: {Top Trends in Workforce Management: How Technology Provides
  Significant Value Managing Your People} (2014),
  \url{http://www.oracle.com/us/products/applications/workforce-management-2706797.pdf},
  (Consulted on 08/03/2018)

\bibitem{nah2004study}
Nah, F.F.H.: A study on tolerable waiting time: how long are web users willing
  to wait? Behaviour \& Information Technology  23(3),  153--163 (2004)

\bibitem{Yang:2014:DRA:2588555.2595631}
Yang, F., Tschetter, E., L{\'e}aut{\'e}, X., Ray, N., Merlino, G., Ganguli, D.:
  Druid: A real-time analytical data store. In: Proceedings of the 2014 ACM
  SIGMOD International Conference on Management of Data. pp. 157--168. SIGMOD
  '14, ACM, New York, NY, USA (2014),
  \url{http://doi.acm.org/10.1145/2588555.2595631}

\end{thebibliography}

\end{document}